\documentclass[11pt]{article}
\usepackage{hyperref}
\pdfoutput=1

\begin{document}
\title{Simulations of droplet dynamics using moving mesh interface tracking with adaptive meshing}
\author{Shaoping Quan\\
        \vspace{6pt}
        Institute of High Performance Computing, Singapore $138632$}

\maketitle

\begin{abstract}
The dynamics of drop(s) has been simulated by the finite volume/moving mesh interface tracking method (MMIT) with adaptive mesh refining and coarsening. In MMIT, the interface is of zero thickness and moves in a Lagrangian fashion. A number of mesh refining and mesh coarsening schemes have been developed to distribute the mesh in an optimized way to achieve computing efficiency and accuracy. Three cases are displayed in this video, which include a droplet pair in tandem impulsively accelerated by a gaseous flow, a droplet moving through a convergent-divergent liquid-filled tube, and a droplet moving toward a sloped solid wall. The fluid dynamics videos of the simulations are to be presented in the Gallery of Fluid Motion, $2011$.
\end{abstract}


The two videos for numerical simulations of droplet(s) dynamics are
\href{http://arxiv.org/src/1010.3067v1/anc/dfd2010-he.mpg}{VideoHighResolution}
and
\href{http://arxiv.org/src/1010.3067v1/anc/dfd2010-le.mpg}{VideoLowResolution}.


\end{document}